\documentclass[letterpaper,11pt]{article}
\usepackage{moriond,epsfig}

\usepackage{geometry}
\usepackage{amsmath}
\usepackage{amssymb}
\usepackage{amsthm}
\usepackage{amsfonts}
\usepackage{newlfont}
\usepackage{graphics}
\usepackage{graphicx}
\usepackage{epsfig}
\usepackage{subfigure}
\usepackage{slashed}
\usepackage{floatflt}

\newcommand{\bb}{\bar{b}}

\begin{document}
\vspace*{3.5cm}
\title{LIGHT HIGGS SEARCHES AT THE LHC USING JET SUBSTRUCTURE}


\author{MATHIEU RUBIN}
\address{LPTHE, UPMC Univ. Paris 6, CNRS, 4 place Jussieu 75252 Paris CEDEX 05, France}

\maketitle\abstracts{
  It is widely believed that searching for a light Higgs boson (with a mass around $120$ GeV) in the $WH$ and $ZH$ channels, where $H\rightarrow b\bb$, will be very challenging at the LHC. These proceedings describe how this channel can be recovered as a promising search channel at high $p_t$ by using a subjet analysis procedure.
}

\section{Motivation}

One of the major goals of the LHC is to discover the Higgs boson, which is the only particle of the Standard Model (SM) that has not been seen yet at colliders. Even if we cannot predict its mass, SM precision electroweak fits seem to favour a low-mass Higgs,\cite{Grunewald:2007pm} and LEP data excluded a mass below $114$ GeV.\cite{Barate:2003sz} That's why we are going to focus on the search for a light Higgs boson at the LHC, assuming its mass is between $115$ GeV and $130$ GeV. \\
Such a search is very difficult because the dominant decay channel into a $b\bb$ pair is swamped by backgrounds. To overcome this issue, experiments usually rely on a combination of several search channels including Higgs decay into $\gamma\gamma$, $ZZ^*$, and $\tau^+\tau^-$. Unfortunately, to be efficient discovery channels, they need a large integrated luminosity.\cite{Aad:2009wy,CMSTDR} Therefore, maybe would it be interesting to examine Higgs boson production in association with a $W$ or a $Z$, where the vector boson decays leptonically, and $H$ decays into $b\bb$? These channels may improve our ability to see a low-mass Higgs boson and they offer an opportunity to measure Higgs couplings with vector bosons and $b\bb$. 

\section{Problems and high-$p_t$ solution}

These channels are important search channels at Tevatron,\cite{:2009pt} but at the LHC they suffer from very large QCD backgrounds, essentially $Wb\bb$, $Zb\bb$ and $t\bar{t}$. The {\sc Atlas} study\cite{AtlasTDR} also highlighted the fact that you need an exquisite control of background shape in order to identify the signal above it.

To get rid of these problems, one possible solution is to look for high-$p_t$ Higgs bosons, $p_t>p_t^{\mbox{\tiny min}}$ where $p_t^{\mbox{\tiny min}}$ is $200$ or $300$ GeV. Of course, by doing so, we only keep a small fraction of the total $VH$ cross-section ($V$ $=$ $W$ or $Z$). But there are several compensating advantages: (1) the ratio signal over background is increased because the $Vb\bb$ cross-section falls somewhat more quickly with $p_t$ than $VH$ cross-section, and due to kinematical constraints\footnote{$b$ and $\bb$ from $t$ and $\bar{t}$ decays must be close to each other, see next section}, the $t\bar{t}$ cross-section\footnote{more precisely the part of the $t\bar{t}$ background that looks like the signal} falls even more quickly with $p_t$; (2) as $V$ and $H$ are very boosted, they mainly decay in the central region of the detector, and therefore we gain in detector acceptance; (3) the $ZH$ channel where $Z\rightarrow \nu\bar{\nu}$ becomes more easily visible because of the large missing $E_T$; (4) backgrounds lose the cut induced shape that was problematic in the {\sc Atlas} study. 

\section{The subjet analysis procedure}

As the Higgs boson is very boosted, the $b$ and $\bb$ from its decay are close to each other.\footnote{\label{AngularSeparation} Their angular separation is roughly $R_{b\bb}\simeq\frac{1}{\sqrt{z(1-z)}}\frac{M_H}{p_t}$, where $z$ and $1-z$ are the energy fractions of the $b$~and~$\bb$, $M_H$ and $p_t$ being respectively the mass and the transverse momentum of the Higgs boson} Therefore, we require $1$ high-$p_t$ jet in the event. Once such a jet is identified, a question immediately arises: is it a Higgs jet or a background QCD jet? Of course, this is common problem for LHC analyses and some solutions have already been found in the boosted regime.\cite{Seymour:1994ca,Butterworth:2002tt,Butterworth:2007ke} All of them exploit the soft divergence of gluon emission in QCD, which implies that there is a high probability to emit a soft gluon. Therefore, if one measures the energy fraction of a splitting, one can make a guess as to whether a gluon was emitted or not. By doing so, we can reduce a large part of the quark-gluon and gluon-gluon splittings, even if gluon splitting into $q\bar{q}$ remains irreducible with this method.

\subsection{The mass drop analysis (MD)}

In our study, we use this idea as a first step to suppress as much background as possible in what we call the Mass Drop Analysis. Its goal is to identify the splitting responsible for the large jet mass, while discarding soft emissions. In order to extract the angular scale $R_{b\bb}$ of this splitting (cf note~\ref{AngularSeparation}), we use the Cambridge/Aachen (C/A) algorithm,\cite{Dokshitzer:1997in,Wobisch:1998wt} which is a sequential recombination algorithm like the $k_t$ except that it is ordered in angle rather than in relative transverse momentum. We go back through the clustering history of the highest-$p_t$ jet using the following procedure:
\begin{enumerate}
  \item Break $j$ into $2$ subjets $j_1$ and $j_2$ (such that $m_{j_1}>m_{j_2}$) by undoing its last stage of clustering.
  \item If ($m_{j_1}<\mu m_j$) and $\min(p_{t_{j1}}^2,p_{t_{j2}}^2)\Delta R^2_{j_1,j_2}>y_{\mbox{\tiny cut}}m_j^2$ then exit the loop.\footnote{$\Delta R_{ij}^2=(y_i-y_j)^2+(\phi_i-\phi_j)^2$}
  \item Otherwise, redefine $j$ to be equal to $j_1$ and go back to step $1$.
\end{enumerate}
Here, $\mu$ and $y_{\mbox{\tiny cut}}$ are $2$ parameters of this procedure that measure respectively the importance of the mass drop and the hardness of the splitting. In practice, they were chosen as $\mu = \frac{2}{3}$ and $y_{\mbox{\tiny cut}} = 0.09$.\\

The final jet $j$ that remains after the end of the loop is considered as our Higgs candidate if both $j_1$ and $j_2$ are $b$-tagged.

\subsection{The filtering analysis}

\begin{figure}[htbp]
  \begin{center}
    \includegraphics[scale=0.62]{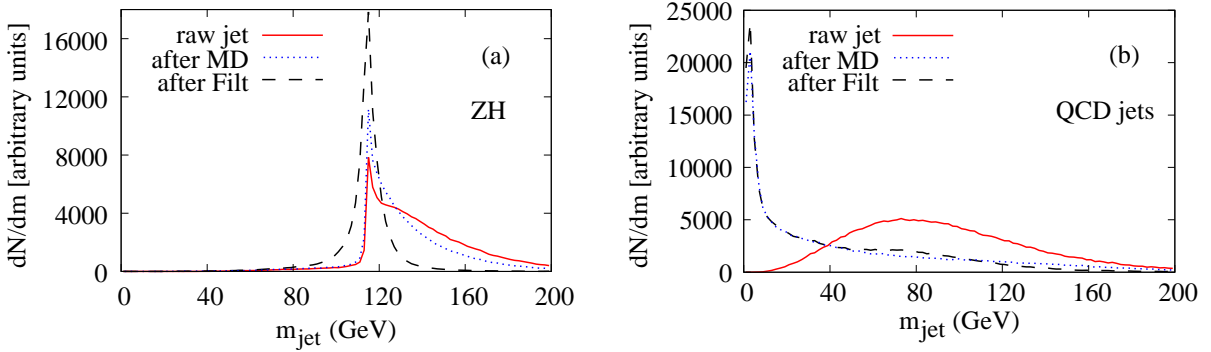}
  \end{center}
  \vspace{-0.2cm}
  \caption[]{(a) Effect of the subjet analysis procedure on the signal: $ZH$ events with UE were generated using HERWIG\cite{Corcella:2002jc,Corcella:2000bw} and JIMMY\cite{Butterworth:1996zw}, with a $p_t$ cut of $200$ GeV, and clustered using C/A with $R=1.2$. We plot the Higgs candidate jet mass distribution before any subjet analysis (solid curve), after MD analysis (dotted curve), and after filtering analysis (dashed curve). (b) The same for the background (QCD jets), where the hardest jet mass distributions are plotted. Notice that the MD analysis essentially reduces the background in the signal region, whereas the filtering analysis improves jet mass resolution for the signal}
  \label{SubjetAnalysisEffect}
\end{figure}

The MD analysis allows us to reduce the background but is not sufficient for the LHC. Indeed, at the reference $p_t$ scale of $200$ or $300$ GeV, the angular separation $R_{b\bb}$ is still large (cf note~\ref{AngularSeparation}: $R_{b\bb}\sim 1$) and the Underlying Event (UE), whose effect on the jet mass scales like $R^4$,\cite{Dasgupta:2007wa} will degrade too much the mass resolution (cf figure~\ref{SubjetAnalysisEffect}). We thus have to reduce this effect as much as possible, and this is what is done in the filtering analysis, whose procedure is the following: first define $R_{\mbox{\tiny filt}}=\min\left(0.3,\frac{R_{b\bb}}{2}\right)$;\footnote{ This choice was motivated by Monte-Carlo studies of a few possible options} then cluster the particles that remain after the MD analysis in the Higgs candidate jet (the ``Higgs neighbourhood'') using C/A algorithm with $R_{\mbox{\tiny filt}}$; and finally take the $3$ hardest jets. This last step allows us to keep the major part of the perturbative radiation, which eventually leads to a good mass resolution on our jets (figure~\ref{SubjetAnalysisEffect}).

\section{Event generation and selection}

To examine the impact of the subjet analysis on a light Higgs boson search at the LHC, we generated signal and background events using HERWIG\cite{Corcella:2002jc,Corcella:2000bw}. We studied $3$ different channels, all involving the production of a Higgs boson (decaying into $b\bb$) with a vector boson (decaying leptonically): ($1$) $ZH$ with $Z\rightarrow l^+l^-$, ($2$) $ZH$ with $Z\rightarrow \nu\bar{\nu}$, ($3$) $WH$ with $W\rightarrow l\nu$. For each channel, we generated several kind of backgrounds: $WW$, $WZ$, $ZZ$, $Zj$, $Wj$, $jj$, $t\bar{t}$, single top. We also simulated the UE using JIMMY\cite{Butterworth:1996zw}.\\
The reader is refered to \cite{MyFirstPaper} in order to get all the detailed cuts of our analysis. But the two most important points to keep in mind are the large $p_t$ cut of at least $200$ GeV we impose on the Higgs candidate, and the subjet analysis procedure to discriminate against QCD background and improve jet mass resolution.

\section{Results at hadron level}

\begin{figure}[hbtp]
  \begin{center}
    \includegraphics[scale=0.24,angle=-90]{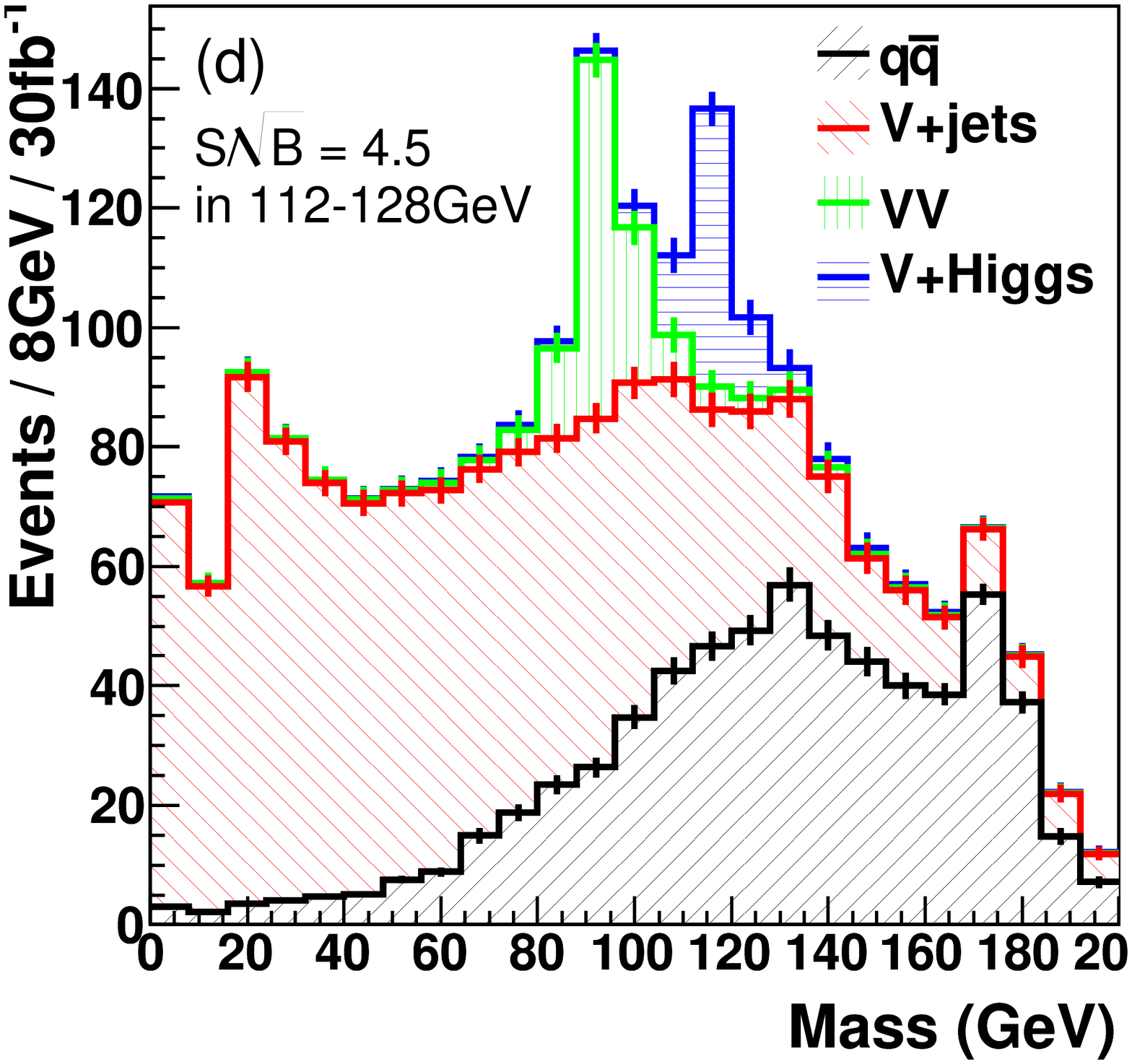}
  \end{center}
  \vspace{-0.2cm}
  \caption{Results for the $3$ channels combined, with $b$-tag/mistag rates = $0.6$/$0.02$}
  \label{results}
\end{figure}
\begin{figure}[hbtp]
  \begin{center}
    \includegraphics[scale=0.43]{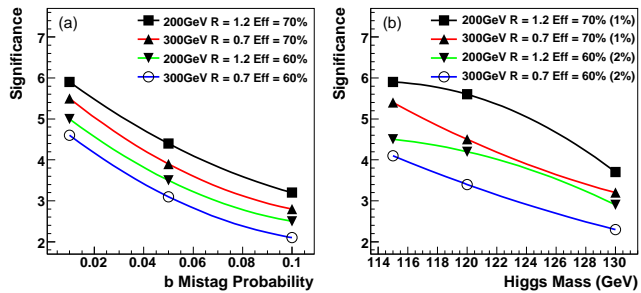}
  \end{center}
  \vspace{-0.2cm}
  \caption{Impact of $b$-tagging, Higgs mass and $p_t$ cut. In (a) $m_H=115$ GeV}
  \label{curves}
\end{figure}

Figure~\ref{results} shows our results for a luminosity of $30\mbox{ fb}^{-1}$ when we combine the $3$ channels. The most important backgrounds are from $t\bar{t}$ and $Vj$. The $VZ$ background can become problematic only in case of poor experimental mass resolution. On the plot, the Higgs peak is seen with a significance of $4.5\sigma$ in a $16$ GeV window ($112$-$128$ GeV).\\
Also shown in figure~\ref{curves} are the results for different $b$-tag/mistag rates and for different Higgs masses.
Most scenarios are above $3\sigma$, but to be a significant discovery channel, it requires decent $b$-tagging, lowish mass Higgs, and good experimental mass resolution.

\section{Conclusion}

Contrary to what was believed, $WH$ and $ZH$ channels may be promising search channels at the LHC for a low-mass Higgs boson. We carried out a simple analysis that leaves room for improvement, and the significance of $4.5\sigma$ that we obtained for $30\mbox{ fb}^{-1}$ implies that it deserves at least a serious experimental study.\footnote{In addition to $b$-tagging, the effects of pile-up, intrinsic resolution and granularity of the detector have to be taken into account} Moreover, this channel provides very important information on $WWH$, $ZZH$ and $b\bb H$ couplings.\cite{Lafaye:2009vr}\\
This study exploits a new subjet analysis procedure that allows one to reduce QCD background and improve mass resolution on jets. It is not restricted to Higgs physics, but can also be used to identify for instance a $W$ or a $Z$, or even new particles, that decay hadronically.

\section*{Acknowledgements}

I would like to thank Jon Butterworth, Adam Davison, and Gavin Salam for collaboration on this work, and the Moriond organizers for such a nice conference!\\






{\bf References}


\begin{thebibliography}{99}

\bibitem{MyFirstPaper}
  J.~M.~Butterworth, A.~R.~Davison, M.~Rubin and G.~P.~Salam,
  Phys.\ Rev.\ Lett.\  {\bf 100}, 242001 (2008)

\bibitem{Grunewald:2007pm}
  M.~W.~Grunewald,
  J.\ Phys.\ Conf.\ Ser.\  {\bf 110} (2008) 042008

\bibitem{Barate:2003sz}
  R.~Barate {\it et al.}  [LEP Working Group for Higgs boson searches],
  Phys.\ Lett.\  B {\bf 565} (2003) 61

\bibitem{Aad:2009wy}
  G.~Aad {\it et al.}  [The ATLAS Collaboration],
  [arXiv:0901.0512]

\bibitem{CMSTDR}
  G.~L.~Bayatian {\it et al.}  [CMS Collaboration],
  J.\ Phys.\ G {\bf 34} (2007) 995.

\bibitem{:2009pt}
  The TEVNPH Working Group [CDF and D0 Collaborations],
  arXiv:0903.4001 [hep-ex].

\bibitem{AtlasTDR}
  The ATLAS Collaboration,
  CERN-LHCC-99-15

\bibitem{Dokshitzer:1997in}
  Y.~L.~Dokshitzer, G.~D.~Leder, S.~Moretti and B.~R.~Webber,
  JHEP {\bf 9708} (1997) 001

\bibitem{Wobisch:1998wt}
  M.~Wobisch and T.~Wengler,
  arXiv:hep-ph/9907280.

\bibitem{Seymour:1994ca}
  M.~H.~Seymour,
  Z.\ Phys.\  C {\bf 63} (1994) 99.

 \bibitem{Butterworth:2002tt}
   J.~M.~Butterworth, B.~E.~Cox and J.~R.~Forshaw,
   Phys.\ Rev.\  D {\bf 65} (2002) 096014

\bibitem{Butterworth:2007ke}
  J.~M.~Butterworth, J.~R.~Ellis and A.~R.~Raklev,
  JHEP {\bf 0705} (2007) 033

\bibitem{Dasgupta:2007wa}
  M.~Dasgupta, L.~Magnea and G.~P.~Salam,
  JHEP {\bf 0802} (2008) 055

\bibitem{Corcella:2002jc}
  G.~Corcella {\it et al.},
  arXiv:hep-ph/0210213.

\bibitem{Corcella:2000bw}
  G.~Corcella {\it et al.},
  JHEP {\bf 0101} (2001) 010

\bibitem{Butterworth:1996zw}
  J.~M.~Butterworth, J.~R.~Forshaw and M.~H.~Seymour,
  Z.\ Phys.\  C {\bf 72} (1996) 637

\bibitem{Lafaye:2009vr}
  R.~Lafaye, T.~Plehn, M.~Rauch, D.~Zerwas and M.~D'uhrssen,
  arXiv:0904.3866 [hep-ph].


\end{thebibliography}

\end{document}